\documentclass[prx,a4paper,aps,twocolumn,superscriptaddress,nofootinbib,10pt]{revtex4-2}

\usepackage{amsmath,amsthm,amsfonts,amssymb,mathrsfs,mathdots,graphicx,xcolor,times,xfrac,mathtools,enumitem,xr,subfigure,bbm,verbatim,comment,dsfont,soul,array,multirow}

\usepackage{comment}
\usepackage[normalem]{ulem}
\usepackage{color}
\definecolor{darkgreen}{RGB}{50,190,50}
\definecolor{darkblue}{RGB}{0,0,190}
\definecolor{darkred}{RGB}{238,0,0}
\definecolor{quantum}{RGB}{83,37,127}
\definecolor{quantumlight}{RGB}{169,146,191}
\definecolor{nice}{RGB}{230,0,230}
\definecolor{nicepink}{rgb}{0.858, 0.188, 0.478}
\usepackage[unicode=true,bookmarks=true,bookmarksnumbered=false,bookmarksopen=false,breaklinks=false,pdfborder={0 0 1}, backref=false,colorlinks=true]{hyperref}
\hypersetup{
	bookmarksnumbered,
	pdfstartview={FitH},
	citecolor={darkgreen},
    linkcolor={darkred},
    linktoc={page},
	urlcolor={darkblue},
	pdfpagemode={UseOutlines}}
\usepackage{tikz}
\usetikzlibrary{shapes,backgrounds,fit,decorations.pathreplacing,arrows,decorations.markings}
\usepackage{verbatim}
\tikzstyle{vecArrow} = [thick, decoration={markings,mark=at position
   1 with {\arrow[semithick]{open triangle 60}}},
   double distance=1.4pt, shorten >= 5.5pt,
   preaction = {decorate},
   postaction = {draw,line width=1.4pt, white,shorten >= 4.5pt}]
\tikzstyle{innerWhite} = [semithick, white,line width=1.4pt, shorten >= 4.5pt]

\hypersetup{
	bookmarksnumbered,
	pdfstartview={FitH},
	citecolor={darkgreen},
	linkcolor={darkred},
	urlcolor={darkblue},
	pdfpagemode={UseOutlines}}
\definecolor{darkgreen}{RGB}{50,190,50}
\definecolor{darkblue}{RGB}{0,0,190}
\definecolor{darkred}{RGB}{238,0,0}
\definecolor{quantum}{RGB}{83,37,127}
\definecolor{quantumlight}{RGB}{169,146,191}
\definecolor{darkorange}{RGB}{255,100,0}
\usepackage{soul}
\usepackage{physics}
\usepackage{bm}
\usepackage{bbm}
\usepackage{bbold}

\usepackage{pifont}

\usepackage{array}

\makeatletter
\theoremstyle{plain}
\newtheorem{thm}{\protect\theoremname}
\theoremstyle{plain}

\theoremstyle{plain}

\theoremstyle{remark}
\newtheorem*{rem*}{\protect\remarkname}
\theoremstyle{plain}

\theoremstyle{plain}

\theoremstyle{definition}

\theoremstyle{plain}
\newtheorem*{thm*}{\protect\theoremname}
\theoremstyle{plain}
\newtheorem*{lem*}{\protect\lemmaname}
\theoremstyle{plain}

\providecommand{\propositionname}{Proposition}
\providecommand{\theoremname}{Theorem}
\providecommand{\lemmaname}{Lemma}
\providecommand{\remarkname}{Remark}
\providecommand{\conjecturename}{Conjecture}
\providecommand{\definitionname}{Definition}
\providecommand{\corollaryname}{Corollary}
\providecommand{\observationname}{Observation}
\allowdisplaybreaks

\def\bra#1{\langle{#1}\vert}
\def\ket#1{\vert{#1}\rangle}
\def\braket#1{\langle{#1}\rangle}

\def\BraVert{e.g.,roup\,\mid\,\bgroup}

\newcommand{\s}[1]{\mathrm{#1}}

\let\oldaddcontentsline\addcontentsline

\newcommand{\starttocentries}{\let\addcontentsline\oldaddcontentsline}
\usepackage[ruled,vlined]{algorithm2e}
\usepackage[mathscr]{eucal}
\usepackage{eucal}

\newcommand{\ot}{\otimes}

\definecolor{plbred}{rgb}{0.8,0,0}

\begin{document}

\title{Fundamental limits on anomalous energy flows in correlated quantum systems}
\author{Patryk Lipka-Bartosik}
\email{patryk.lipka.bartosik@gmail.com}
\affiliation{Department of Applied Physics, University of Geneva, 1211 Geneva, Switzerland}

\author{Giovanni Francesco Diotallevi}
\affiliation{Augsburg University$,$ Institute of Physics$,$ Universitätsstraße 1 (Physik Nord)$,$ 86159 Augsburg}
\affiliation{Department of Physics and Nanolund, Lund University, Box 118, 221 00 Lund, Sweden}

\author{Pharnam Bakhshinezhad}
\email{pharnam.bakhshinezhad@tuwien.ac.at}
\thanks{Formerly known as Faraj Bakhshinezhad}
\affiliation{Atominstitut, Technische Universit{\"a}t Wien, Stadionallee 2, 1020 Vienna, Austria}
\affiliation{Department of Physics and Nanolund, Lund University, Box 118, 221 00 Lund, Sweden}

\date{\today}

\begin{abstract} 
In classical thermodynamics energy always flows from the hotter system to the colder one. However, if these systems are initially correlated, the energy flow can reverse, making the cold system colder and the hot system hotter. This intriguing phenomenon is called ``anomalous energy flow'' and shows the importance of initial correlations in determining physical properties of thermodynamic systems. Here we investigate the fundamental limits of this effect. Specifically, we find the optimal amount of energy that can be transferred between quantum systems under closed and reversible dynamics, which then allows us to characterize the anomalous energy flow. We then explore a more general scenario where the energy flow is mediated by an ancillary quantum system that acts as a catalyst. We show that this approach allows for exploiting previously inaccessible types of correlations, ultimately resulting in an energy transfer that surpasses our fundamental bound. To demonstrate these findings, we use a well-studied quantum optics setup involving two atoms coupled to an optical cavity.
\end{abstract}

\keywords{}

\maketitle

\section{Introduction}

The second law of thermodynamics, as formulated by Clausius, states that heat always flows spontaneously from a hot to a cold system \cite{clausius1867mechanical}. The validity of this formulation relies on the assumption that the two systems share no correlations~\cite{boltzmann2003relation,lebowitz1993boltzmann,zeh1989direction}. When this condition is not met, various counter-intuitive phenomena can occur \cite{Lloyd1989,hakim1985quantum,haake1985strong,rio2011thermodynamic,partovi2008entanglement,rivas2014quantum,sapienza2019correlations,breuer2016colloquium,Hartman2004,hernandez2015locality,Kliesch2014,ferraro2012intensive,romero1997decoherence,Campisi2009}. This suggest that classical thermodynamic concepts, such as work or heat, may need to be appropriately modified before they can be applied to correlated quantum systems \cite{beraEtAl2017,AlipourSciRep_2016,AlipourArxiv2021,PhysRevA.105.022204,AfsaryPRA2020}. Indeed, it is now well-established that initial correlations can influence the work performed in a thermodynamic process \cite{huber2015thermodynamic,Perarnau2015,Allahverdyan2000,Ford2006,Bruschi2015,Friis2016,Salvia_2022,BakhshinezhadJPA_2019}, or enhance the performance of certain thermal machines  \cite{doyeux2016quantum,PhysRevA.91.012117,hewgill2018quantum,latuneEtAl2021,HolsworthPhysRevA2022}.

In this work we examine how initial correlations affect energy transfer between quantum systems. This problem was raised in Ref.~\cite{partovi_2008}, and has since then been explored both theoretically \cite{jevtic2012maximally, ahmadi_salimi_khorashad_2021, Jennings_2010,Colla_2022,Mirza2023,Zicari2020,modi2011preparation,dijkstra2010non, smirne2010initial,chaudhry2013role,semin2012initial,dajka2010distance,zhang2010different,zhang2015role,PhysRevResearch.3.023235} and experimentally \cite{micadei2019}. These studies have shown that in the presence of initial correlations, energy transfer can be amplified or even reversed. Our aim is to understand the fundamental limitations of this process.


Let us examine a closed quantum system comprising two locally thermal and non-interacting parts $\s{A}$ and $\s{B}$ with inverse temperatures $\beta_\s{A}$ and $\beta_{\s{B}} \leq \beta_{\s{A}}$. Quantum mechanics imposes limits on how much energy can be transferred between such systems \cite{Jennings_2010}. Specifically, in any unitary process, the energy change of the colder system, denoted $\Delta E_{\s{A}}$, satisfies
\begin{align}
    \label{eq:main_bound}
     \Delta E_{\s{A}}\left(\beta_{\s{A}} - \beta_{\s{B}}\right) \geqslant \Delta \mathcal{I} -\,\beta_{\s{B}}W,
\end{align}
where $W := \Delta E_{\s{A}} + \Delta E_{\s{B}}$ is the external work performed on the bipartite system, and $\Delta \mathcal{I}$ is the change in the quantum mutual information between $\s{A}$ and $\s{B}$ during the process, see Appendix \ref{app:bound on REF} for more details.

\begin{figure}[h]
    \centering
    \includegraphics[width=\linewidth]{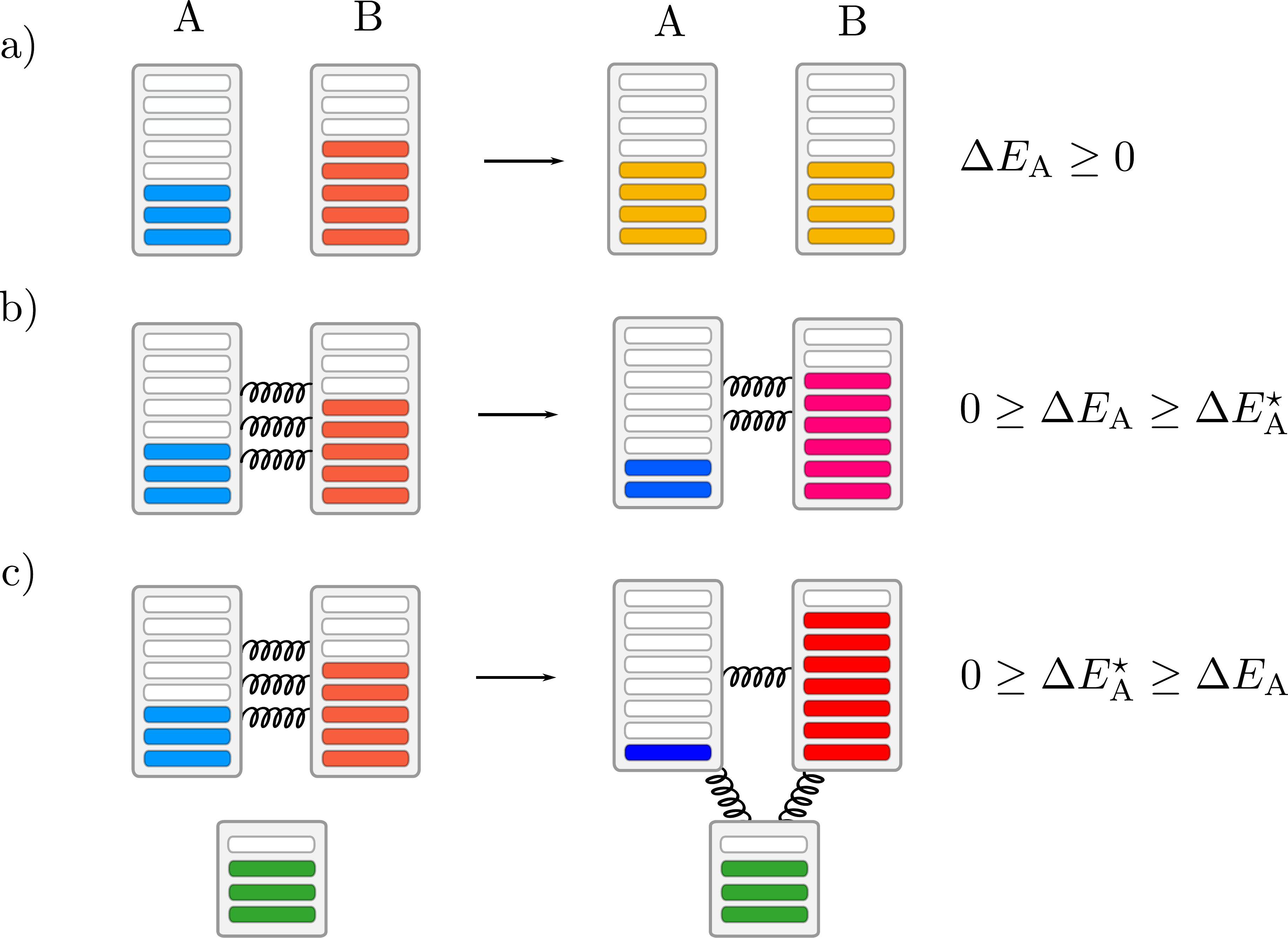}
    \caption{
    \textbf{Energy flows in quantum systems.} Panel $(a)$ shows the standard scenario of two initially uncorrelated systems evolving unitarily. In this case, in the absence of work, energy always flows from the hot $(\s{B})$ to the cold $(\s{A})$ system, hence the local energy change $\Delta E_{\s{A}} \geq 0$.  In Panel $(b)$ the two systems are initially correlated and an anomalous energy flow can be observed, i.e. $\Delta E_{\s{A}} < 0$. In this work we determine the minimal value of $\Delta E_{\s{A}}$ optimized over energy-preserving and unitary evolutions, denoted with $\Delta E_{\s{A}}^{\star} \leq 0$. Panel $(c)$ shows the extended scenario involving a quantum catalyst $(\s{C})$, i.e. an ancillary system that returns to its initial state at the end of the process. The catalyst further enhances the energy flow, i.e. leads to  $\Delta E_{\s{A}} < \Delta E_{\s{A}}^{\star}$, and can be later reused for another process.
    }
    \label{fig:example}
\end{figure}

In classical thermodynamics, the two systems are always initially uncorrelated, i.e. $\Delta \mathcal{I} \geq 0$. Therefore, lowering the energy of the cold system $\s{A}$ can only be accomplished by performing work on the system, i.e. $W> 0$. On the other hand, when no work is performed, then $\Delta E_{\s{A}}$ will have the same sign as $\beta_{\s{A}} - \beta_{\s{B}}$  (i.e., heat flows from hot to cold). Consequently, when $\Delta \mathcal{I} \geq 0$, Eq. (\ref{eq:main_bound}) reduces to the Clausius' statement of the second law of thermodynamics: ``It is impossible to transfer energy from a cold system to a hot one in a cyclic process without performing work''~\cite{clausius1867mechanical}. In this sense Eq. (\ref{eq:main_bound}) can be viewed as a generalization of the Clausius formulation of the second law of thermodynamics which takes into account initial correlations between subsystems.

This natural flow of energy can be violated if subsystems are initially correlated \emph{and} the process consumes correlations, which corresponds to $\Delta \mathcal{I} < 0$  and $W = 0$ in Eq. (\ref{eq:main_bound}). Under this conditions, quantum mechanics permits processes that transfer energy from the cold system to the hot one, i.e. for $\Delta E_{\s{A}} < 0$. 
This phenomenon is known as \emph{anomalous energy flow} (AEF) and demonstrates that correlations can act as a thermodynamic resource. 

One can also consider other types of resources to accomplish or improve thermodynamic tasks: the available level of control on the system and its energy structure. These concepts, known as control and structural complexities, are crucial in the context of AEF \cite{Landauer_vs_Nernst}. Indeed, restricting the allowed level of control constrains the possible evolution of the system.  Enlarging the set of allowed operations usually relaxes these constraints. For that, we will introduce an ancillary system (a catalyst) that assists the process by mediating the flow of energy between different parts of the system without providing energy itself. This extension will allow us to investigate the impact of control and energy structure on energy flows in quantum systems.

Here we investigate the fundamental limits on energy flows in out-of-equilibrium quantum systems. For a general bipartite quantum state $\varrho_{\s{AB}}$, we determine the maximal amount of energy that can be transferred between $\s{A}$ and $\s{B}$ via energy-conserving unitaries. In the special case where the local states of $\varrho_{\s{AB}}$ are thermal with inverse temperatures $\beta_{\s{A}} > \beta_{\s{B}}$, this leads to the optimal bound on AEF. We then discuss a way to overcome this limitation by using \emph{coherent quantum catalysis} \cite{Lipka_Bartosik_2023}. We demonstrate these findings using an experimentally-relevant model in quantum optics. 

\section{Framework}
We consider a closed quantum system composed of two subsystems $\s{A}$ and $\s{B}$. Each subsystem comes with a local Hamiltonian $H_x = \sum_{i=1}^{d_x}\varepsilon_{i}^{x} \ket{i}\bra{i}_x$ of dimension $d_x$, where $\varepsilon_{i}^x$ indicates the $i$-th energy eigenvalue and $\ket{i}_x$ the corresponding eigenstate of subsystem $x\, \in\,\{\, \s{A},\, \s{B}\}$. Without loss of generality, we assume that the energies are sorted in an increasing order, i.e. $\varepsilon_{i}^x \,\geq\,\varepsilon_{j}^x $ for all  $i\geq j$. 
We further denote the free Hamiltonian of $\s{AB}$ with $H_0\, =\, \sum_{\nu=1}^{M}\mathcal{E}_{\nu}\, \Pi_{\nu}$, where $M$ is the number of distinct energy levels $\{\mathcal{E}_{\nu}\}$ of the total system and $\{\Pi_{\nu}\}$  are projectors onto the subspaces corresponding to the $(m_\nu)-$fold degenerate eigenvalue $\mathcal{E}_{\nu}$, so that $\sum_{\nu=1}^M m_{\nu} = d_{\s{A}} d_{\s{B}}$. The average energy of subsystem $x$ in the state $\varrho_x$ is denoted with $E(\varrho_x)\,:=\,\Tr[H_x \, \varrho_x]$. 

We are interested in driving the bipartite system $\s{AB}$ unitarily to study the transfer of energy from $\s{A}$ to $\s{B}$. This process is modelled using a Hamiltonian $H_{\s{AB}}(t)\,=\,H_0\,+ \, V(t)$, where $H_0 \,=\,H_{\s{A}}\,+\,H_{\s{B}}$ and $V(t)$ is a cyclic potential. As a result, the system evolves as $\varrho_{\s{AB}}\rightarrow {\varrho_{\s{AB}}^\prime}=
    U\, \varrho_{\s{AB}}\,U^{\dagger}$, where $U\,=\, \mathcal{T}[\s{exp}(-i\int_{0}^{\tau} H_{\s{AB}}(t) \mathrm{d}t )]$ is a unitary operator, $\tau$ is the duration of the protocol, and ${\mathcal{T}}$ is the time-ordering superoperator (we use $\hbar= \, k_{\s{B}}=1$). Importantly, any unitary operation $U$ can be implemented in the above manner. Notice that the initial and the final states $\varrho_{\s{AB}}$ and $\varrho_{\s{AB}}'$ will, in general, not be of the Gibbs form.
    
    It is often the case that fundamental (or experimental) limitations on the cyclic potential $V(t)$ restrict the class of unitaries that can be implemented. For instance, in the absence of work, the total energy of the system is conserved at all times $t \in [0, \tau]$. This can be captured mathematically by requiring that $[H_{0}, V(t)]=0$ at all times $t$ or, equivalently, that $H_0$ and $U$ share the same eigenbasis, i.e. $[H_0, U] = 0.$ In other words, the unitary $U$ has a block-diagonal form
\begin{equation}
    U\,=\, \prod_{\nu=1}^{M} U_{\nu},\label{eq: block diagonal preserving unit}
\end{equation} 
where $U_{\nu}$ acts non-trivially only on the subspace $\nu$.

An example scenario in which energy conservation imposes dynamical limitations can be observed when A and B represent two two-level systems. In this setting, the two systems are unable to interact with each other, unless they shared an identical energy gap.  However, if they were out of resonance, one could still bring an ancillary system capable of providing the missing energy, thus enabling non-trivial interactions. Such an ancilla, however, must recover its energy back by the end of the protocol, otherwise it would itself contribute to the flow of energy. Hence, in Sec. \ref{subsec:coherent_cat} we extend the framework by allowing for using ancillary systems which must be left unchanged at the end of the protocol. In classical thermodynamics the use of such systems is usually neglected (e.g. the piston used to compress a gas). Such systems can be mathematically modelled by adding an ancillary subsystem (catalyst) ${\s{C}}$ with a Hamiltonian $H_{\s{C}}$ which interacts with the system ${\s{AB}}$, and after the interaction returns to its initial state. In this sense the joint system $\s{ABC}$ evolves according to an energy-preserving unitary $U$ as $\varrho_{\s{AB}} \otimes \omega_{\s{C}} \rightarrow \varrho_{\s{ABC}}' := U(\varrho_{\s{AB}} \otimes \omega_{\s{C}})U^{\dagger}$ under the constraint that $\Tr_{\s{AB}}[\varrho'_{\s{ABC}}] = \omega_{\s{C}}$. Quantum catalysis is an active topic of research which finds applications in quantum thermodynamics \cite{Brand_o_2015,Ng_2015,Wilming2017,Mueller2018,Lipka-Bartosik2021,shiraishi2021quantum,Gallego2016,boes2019bypassing,Boes2019,Henao_2021,Henao2022,son2022catalysis,Lipka_Bartosik_2023,Seok2023}, entanglement theory \cite{Jonathan_1999,vanDam2003,turgut2007catalytic,daftuar2001mathematical,sun2005existence,feng2005catalyst,PhysRevLett.127.080502,Kondra_2021,Datta2022}, and other fields \cite{Aberg2014,Vaccaro2018,Marvian2019,lostaglio2019coherence,takagi2022correlation,wilming2021entropy,char2023catalytic,rubboli2022fundamental,van2023covariant,lie2021catalytic}. See Refs. \cite{datta2022catalysis,lipkabartosik2023catalysis} for recent reviews.

Our main figure of merit is the maximal energy flow between systems $(\s{A})$ and $(\s{B})$, denoted by $\Delta E^{\star}_{\s{A}}$, i.e.
\begin{align}
    \label{eq:energy_flow_rev}
    \Delta E^{\star}_{\s{A}} := \min_{U} \left[ E(\sigma_\s{A}) - E(\varrho_\s{A})\right],
\end{align}
where $\sigma_\s{A} := \Tr_{\s{B}}\left[U\varrho_\s{AB} U^{\dagger}\right]$ and $U$ is a unitary that satisfies $[U, H_\s{AB}] = 0$. We will also be interested in the special case when initial state $\varrho_{\s{AB}}$ has thermal marginals, i.e. $\Tr_{\s{A}}[\varrho_{\s{AB}}] =  \gamma_{\s{B}}$ and $\Tr_{\s{B}}[\varrho_{\s{AB}}] = \gamma_{\s{A}}$, where $\gamma_x = \s{exp}(-\beta_xH_x)/\mathcal{Z}_x$ is a thermal state of subsystem $x \in\{\s{A},\s{B}\}$ with $\mathcal{Z}_x = \Tr[\s{exp}(-\beta_x H_x)]$, and the inverse temperatures satisfy $\beta_{\s{A}}\geq \beta_{\s{B}}$. In this context the quantity $\Delta E_{\s{A}}^*$ captures the maximal degree of reversing the natural flow of energy between subsystems $\s{A}$ and $\s{B}$, i.e. the \emph{anomalous energy flow} (AEF). In this case $\Delta E_{\s{A}}^{\star} < 0$ indicates that AEF can occur for some unitary process $U$, and $\Delta E_{\s{A}}^{\star} \geq 0$ means that the energy always flows from $\s{A}$ to $\s{B}$, and hence AEF cannot occur. 


\section{Results}

\subsection{Optimal energy flow in a correlated quantum system}

Our first contribution is finding the maximal energy flow $\Delta E_{\s{A}}^{\star}$ as defined via Eq. (\ref{eq:energy_flow_rev}) for general quantum states {$\varrho_{\s{AB}}$}. We start by observing that energy-preserving unitaries simply rotate populations within degenerate energy eigenspaces of the joint system $\s{AB}$. Let us therefore decompose $\varrho_\s{AB}$ as
\begin{align}
    \varrho_{\s{AB}} &= \sum_{\nu = 1}^M \varrho_{\s{AB}}^{\nu\nu} + \sum_{\nu \neq \mu}^M \varrho_{\s{AB}}^{\nu\mu},
   \label{eq: Blocky density operator}
    \end{align} 
where $\varrho_{\s{AB}}^{\nu \mu}:=\, \Pi_\nu\,\varrho_{\s{AB}} \, \Pi_\mu $. 
Using this representation, in Appendix \ref{sec: use of nondegenerat coherences} we show that the operators $\varrho^{\nu\mu}_{\s{AB}}$ for $\nu \neq \mu$ (i.e., elements outside of degenerate energy subspaces) do not contribute to the average local energy $E(\varrho_{\s{A}})$. Consequently, energy coherences outside of the degenerate energy subspaces (and hence also the correlations they describe), are irrelevant from the perspective of changing local energies of the system. Therefore, the only relevant contribution towards $\Delta E_{\s{A}}$ comes from the operators $\varrho_{\s{AB}}^{\nu \nu}$. As these operators are independent from each other, optimizing local energy change on $\s{A}$ can be achieved by optimizing it within each subspace $\nu$.  

Let us now observe that the unitary operators acting on subspace $\nu$, i.e. $u_\nu=\Pi_\nu U_{\nu}$, can be arbitrary. This means that in each such subspace, we can apply a different {$u_{\nu}$} that optimizes energy flow within that subspace, e.g. using the techniques described in Ref. \cite{Exponential_Improve_for_Quant_Cool}. More specifically, let us write $\varrho_{\s{AB}}^{\nu \nu} = \sum_{n=1}^{m_{\nu}} p_{n}^{\nu {\downarrow}} \ketbra{\phi^{\nu}_n}$, where $p^{\nu{\downarrow}}_{n}$ are the eigenvalues ordered non-decreasingly in the degenerate subspace $\nu$ and $\ket{\phi^{\nu}_{n}}$ are the corresponding eigenvectors. Note that $x^\downarrow_n$ indicates the elements of a set ordered in a non-decreasing order. Moreover, we also decompose $\Pi_{\nu}\, =\, \sum_{n=1}^{m_\nu} \, \ketbra{\mathcal{E}^{\nu}_{n}}$, where $\ket{\mathcal{E}^{\nu}_{n}}$ are sorted in the order of non-increasing local energies of $\s{A}$, i.e. such that $\braket{\mathcal{E}_{n}^{\nu}|H_{\s{A}} \ot \mathbb{1}_{\s{B}}|\mathcal{E}_{n}^{\nu}} \geq \braket{\mathcal{E}_{k}^{\nu}|H_{\s{A}} \ot \mathbb{1}_{\s{B}}|\mathcal{E}_{k}^{\nu}}$ for $n \geq k$. In this notation, the unitary $u_{\nu}$ that minimizes the local energy of subsystem $\s{A}$ within the degenerate energy subspace $\nu$ is given by $u_{\nu} = \sum_{n = 1}^{m_{\nu}} \ket{\mathcal{E}^{\nu}_{n}}\bra{\phi^{\nu\downarrow}_{n}}$. In other words, in each subspace $\nu$ we apply a unitary which has the following two actions: $(1)$ diagonalize $\varrho^{\nu\nu}_{\s{AB}}$, and $(2)$ permute the eigenvalues of $\varrho^{\nu\nu}_{\s{AB}}$ in the order of non-decreasing local energies of $\s{A}$.  For more details see Appendix \ref{app:proof_theorem1}.

The above reasoning can be formulated in terms of the following theorem, which constitutes our first main result. 
\begin{thm}
\label{thm1}
For a bipartite state $\varrho_{\s{AB}}$ the optimal local energy change under energy-preserving unitaries is given by
\begin{align}
    \Delta E_{\s{A}}^{\star} = E(\sigma_{\s{A}}^{\star}) - E(\varrho_{\s{A}}),
\end{align}
where $\sigma_{\s{A}}^{\star} = \sum_{i=1}^{d_\s{A}} q_{i}^{\star}\ketbra{\varepsilon_{i}^{\s{A}}}$ and $q_{i}^{\star}$ are given by 
\begin{align}
    q_{i}^{\star} = \sum_{\nu=1}^M\sum_{n=1}^{m_{\nu}}p_n^{\nu {\downarrow}} \braket{\mathcal{E}_n^{\nu}| \left(\dyad{\varepsilon_i^{\s{A}}}\ot \mathbb{1}_{\s{B}}\right)|\mathcal{E}_n^{\nu}}.
\end{align}
The unitary $U_{\star}$ that enables this energy flow is given by
\begin{align}
    \label{eq:opt_arb_unitary}
    U_{\star} = \prod_{\nu=1}^M U_{\nu}, \,\,\, \s{where} \,\, U_{\nu} = \sum_{n=1}^{m_\nu}\ket{\mathcal{E}^{\nu}_{n}}\bra{\phi^{\nu\downarrow}_{n}}+{\sum_{\mu\neq\nu}\Pi_\mu}.
\end{align}
\end{thm}

The above theorem characterizes the optimal reversible (unitary) and energy-preserving protocol which minimizes the (local) energy change on subsystem $\s{A}$. It therefore provides the solution to the optimization problem from Eq. (\ref{eq:energy_flow_rev}). In particular, when $\Delta E_{\s{A}}^{\star} < 0$ and local marginals of {$\varrho_{\s{AB}}$} are thermal states with $\beta_{\s{A}} \geq \beta_{\s{B}}$, Theorem \ref{thm1} gives the optimal (achievable) bound on the AEF. In general, the final state obtained in the optimal protocol from Theorem \ref{thm1} is not  thermal.

One of the insights of Theorem \ref{thm1} is that the only type of correlations which can influence the energy flow between $\s{A}$ and $\s{B}$ correspond to degenerate energy subspaces. As a consequence, the bound from Eq. (\ref{eq:main_bound}) for $W = 0$ is not generally achievable with energy-conserving unitaries. Hence, for quantum states that exhibit correlations between non-degenerate energy levels, the natural flow of energy can never be reversed, even though they can be highly correlated. 

Our second main result provides a way to overcome this limitation using the concept of coherent quantum catalysis \cite{Lipka_Bartosik_2023}. This provides a way to access the correlations ``locked'' in the non-degenerate energy subspaces. In other words, we will use a catalyst to lift some of the dynamical restrictions imposed by the energy conservation law \cite{Aberg2014}. In the next section, we show how to surpass the bound given by Theorem \ref{thm1} using coherent quantum catalysis. 

\subsection{Enhancement of energy flow using coherent catalysis}
\label{subsec:coherent_cat}
We start with an experimentally-relevant example of spins coupled to a single-mode optical cavity. Consider two spins (two-level systems) $\s{A}$ and $\s{B}$ with the same energy $\varepsilon$ coupled to  a single mode of an electromagnetic field $\s{C}$ with frequency $\varepsilon$ and a bosonic annihilation operator $a$. Furthermore, let $\sigma_i = \dyad{g}{e}_{i}$ denote the lowering operator of the $i$-th qubit. The systems evolve via the Tavis-Cummings Hamiltonian \cite{dicke1954coherence,tavis1968exact}, which in the
rotating wave approximation reads
\begin{align}
    H 
    = \varepsilon  \sum_{i = \s{A}, \s{B}} \sigma_i^{\dagger} \sigma_i +\varepsilon a^\dagger a + H_{\text{int}}, \label{eq: RW approx hamiltonian}
\end{align}
where $H_{\text{int}} :=  g \sum_{i=\s{A},\s{B}} \left(a \sigma_i^{\dagger} + a^\dagger \sigma_i\right)$ and $[H, H_{\text{int}}] = 0$. The two qubits should be understood as the cold $(\s{A})$ and the hot $(\s{B})$ subsystems. The cavity $\s{C}$ plays the role of the catalyst. The system $\s{AB}$ is prepared in a state $\varrho_{\s{AB}}(\lambda, \theta)$ which is a linear combination of a product of Gibbs states and two orthogonal Bell states, i.e.

\begin{figure}[]
    \centering
    \includegraphics[width=\linewidth]{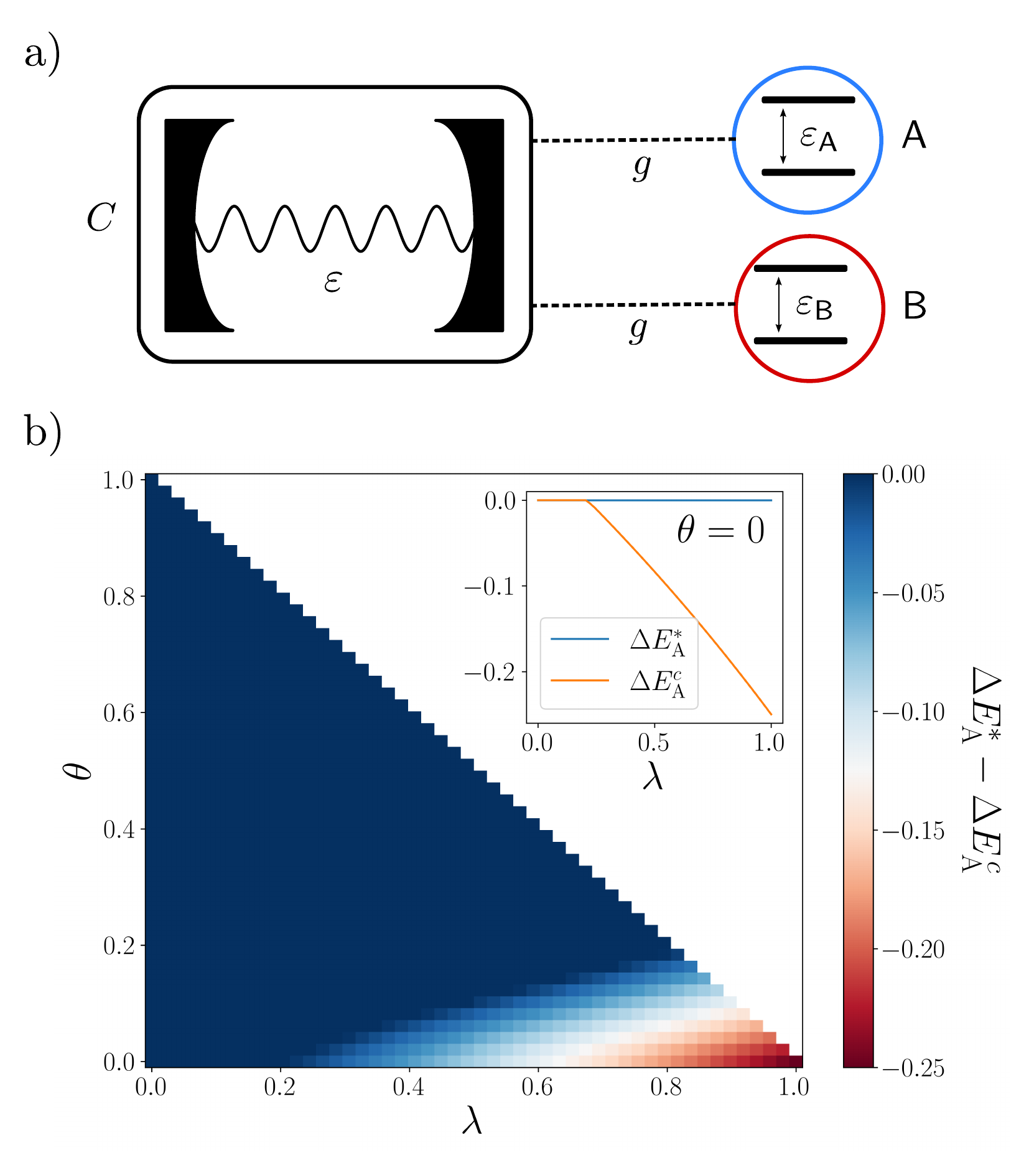}
    \caption{
    \textbf{Catalytic anomalous energy flow.} We consider two spins $\s{A}$ and $\s{B}$ coupled to an optical cavity $\s{C}$ with $N = 3$ Fock levels. Fig. (a) shows the schematic depiction of the Tavis-Cummings model in which two two-level systems are each coupled to a single mode of light in an optical cavity (see main text). Fig (b) shows $\Delta E_{\s{A}}^* - \Delta E_{\s{A}}^c$, i.e. the difference between the optimal energy change on the cold system $\s{A}$ and the energy change obtained using the catalytic protocol, for the state given by Eq. (\ref{eq:state_param}) as a function of parameters $\lambda$ and $\theta$. The inset shows the particular case of $\theta = 0$, for which no AEF is possible due to Theorem $1$ (blue curve). We observe that using the cavity as a catalyst allows for energy flows that would otherwise be impossible to achieve.} 
    \label{fig:example}
\end{figure}

\begin{align}
    \label{eq:state_param}
    \varrho_{\s{AB}}(\lambda, \theta) = (1-\lambda-\theta) \gamma_{\s{A}} \ot \gamma_{\s{B}} + \lambda \phi^+_{\s{AB}} + \theta \psi_{\s{AB}}^-, 
\end{align}
where $\gamma_{i} := \gamma(\beta_{i}, H_{i})$ for $i=\s{A},\s{B}$ and the (inverse) temperatures satisfy $\beta_{\s{A}} > \beta_{\s{B}}$ (i.e. $\s{A}$ is colder than $\s{B}$). The two Bell states are given by $\ket{\phi_{\mathrm{AB}}^{+}} := (\ket{00}_{\s{AB}}+\ket{11}_{\s{AB}})/\sqrt{2}$ and $\ket{\psi_{\s{AB}}^-} := (\ket{01}_{\s{AB}}-\ket{10}_{\s{AB}})/\sqrt{2}$, where $\phi^+_{\s{AB}} := \ketbra{\phi_{\mathrm{AB}}^{+}}$ and $\psi^-_{\s{AB}} := \ketbra{\psi_{\mathrm{AB}}^{-}}$, and the free parameters satisfy $\lambda + \theta \leq 1$. The optical cavity $\s{C}$ will be prepared in a state $\omega_{\s{C}}$ which will be determined later. 

We consider an energy-conserving unitary protocol composed of two steps. First, the spins interact with the cavity via unitary $U(\tau) = \s{exp} \left(-i \int_{0}^\tau H(t)\, \s{d} t\right)$, which leads to
    \begin{align}
        \sigma_{\s{AB}}{(\tau)} = \Tr_{\s{C}}\left[U(\tau)\left(\varrho_{\s{AB}} \ot \omega_{\s{C}} \right) U^{\dagger}(\tau)\right].
    \end{align}
The state of the cavity $\omega_{\s{C}}$ is the unique solution to
\begin{align}
    \label{eq:catalytic_eq}
    \omega_{\s{C}} = \Tr_{\s{AB}}\left[U(\tau)\left(\varrho_{\s{AB}} \ot \omega_{\s{C}} \right) U^{\dagger}(\tau)\right].
\end{align}
In other words, the cavity returns to its initial state at $t = \tau$. This guarantees that $\s{C}$ acts as a catalyst, i.e. its quantum state does not change under the dynamics. Consequently, it delivers no energy to the bipartite system $\s{AB}$ and the energy flow can only be attributed to the system $\s{AB}$. This is valid for any value of the parameter $\tau > 0$. Second, we apply the optimal unitary $U_{\star}$ from Theorem \ref{thm1} to $\sigma_{\s{AB}}(\tau)$, which leads to $\sigma'_{\s{AB}}{(\tau)} := U_{\star} \,\sigma_\s{AB}{(\tau)} \,U_{\star}^{\dagger}$, and the protocol terminates. 

The resulting energy change of subsystem $\s{A}$ is given by $\Delta E_{\s A}(\tau) = E\left(\sigma'_{\s{A}}{(\tau)}\right) - E(\varrho_{\s{A}})$, where $\sigma'_{\s{A}}{(\tau)} = \Tr_{\s{B}} \left[\sigma'_{\s{AB}}{(\tau)}\right]$. The value of parameter $\tau$ is then numerically optimized so that $\Delta E_{\s{A}}(\tau)$ achieves minimal value, which we denote with $\Delta E_{\s{A}}^{c} := \min_{\tau > 0} \Delta E_{\s{A}}(\tau)$. 

In Fig. \ref{fig:example} we present the performance of the above protocol in the context of AEF. More specifically, we observe that $\Delta E_{\s{A}}^{c} $ can be significantly lower than $\Delta E_\s{A}^*$ as determined from Theorem \ref{thm1}. This shows that the use of a (coherent) catalyst can improve the transfer of energy between two subsystems. Moreover, the largest advantage is obtained when $\varrho_{\s{AB}}(\lambda, \theta)$ does not contain ocuppations, nor coherences, in the subspace of degenerate energy, i.e. the one spanned by energy eigenstates $\ket{01}_{\s{AB}}$ and $\ket{10}_{\s{AB}}$ (which corresponds to $\theta = 0$). This leads to a genuinely new type of AEF which relies on the correlations between \emph{non-degenerate} energy levels, as opposed to correlations within the degenerate energy subspace. Moreover, since in the case when $\theta = 0$ we have $\Delta E_{\s{A}}^{\star} = 0$ and thus no AEF can occur, the above protocol demonstrates that coherent catalysis can not only enhance, but also \emph{activate} previously impossible AEFs. It is worth mentioning that the setup we have studied can be experimentally realized using current state-of-the art technology \cite{baumann2010dicke,klinder2015dynamical,zhiqiang2017nonequilibrium,RevModPhys.85.553}.


The example presented above demonstrates that AEF  can be enhanced, and even activated, using appropriately tuned quantum states acting as catalysts. It is natural to ask about the optimal energy flow that can be obtained when using arbitrary catalysts and energy-conserving interactions. In Appendix \ref{app:bound_on_REF} we derive such a fundamental bound which sets a lower bound on the value of $\Delta E$ that can be achieved using a catalyst. The bound is given by a semi-definite program (SDP) \cite{vandenberghe1996semidefinite,Skrzypczyk2023}, which guarantees that it can be efficiently solved using existing numerical techniques \cite{cvx}. Finally, in Appendix \ref{app:toy_example} we present a toy example which shows that the activation of AEF can also occur when the system $\varrho_{\s{AB}}$ contains purely classical correlations. In this case the catalyst required by the transformation is also fully classical (i.e. incoherent in the energy basis). 


\section{Summary}
We investigated the role of correlations in reversing the natural flow of energy in correlated quantum systems. More specifically, we derived the maximal (local) energy change that can be achieved using energy-conserving unitary processes, and determined the actual process that achieves this. This allows to determine the optimal anomalous energy flow (AEF) that can be observed in correlated quantum systems. Furthermore, we also investigated this task in the presence of a quantum catalyst and shown that AEF can be further enhanced, and even activated, when an appropriate ancillary system is used catalytically. 

In this work we addressed the fundamental question of how correlations can help in transferring energy between two quantum systems. However, the complete understanding of this problem is far from being achieved. Here we assumed that the unitary can be implemented perfectly by having access to ideal interactions and ideal time-keeping, which has an infinite energy cost. In realistic (imperfect) scenarios, more restrictions are put in the accessible operations, which can, in principle, reduce the amount of energy that can be transferred between the two systems. It would be interesting then to see how the presence of an imperfect clock and control affect our results. Furthermore, it would be interesting to see whether coherent catalysis can be used to improve actual thermodynamic protocols. For example, our results suggest that coherent catalysis not only allows to generate a new kind of AEF, but that it can also be used to improve the existing optimal protocols for cooling or work extraction by explicitly exploiting the energy coherence (and entanglement) of quantum states.  


\section{Acknowledgements} We are grateful to S. H. Lie for multiple insightful comments on the first draft of this work, and in particular for spotting a gap in the proof of Theorem $1$ and proposing an alternative proof. We further thank P. Samuelsson, P. P. Potts, B. Annby-Andersson, G. Vitagliano,  {M.} Perarnau-Llobet and {R.} Silva for fruitful discussions.  PLB acknowledges the Swiss National Science Foundation for financial support through the NCCR SwissMAP. GFD acknowledges the Wallenberg Center for Quantum Technology (WACQT) for financial support via the EDU-WACQT program funded by Marianne and Marcus Wallenberg Foundation. PB is supported by grant number FQXi Grant Number: FQXi-IAF19-07 from the Foundational Questions Institute Fund, a donor advised fund of Silicon Valley Community Foundation and funding from the European Research Council (Consolidator grant ‘Cocoquest’ 101043705).

\appendix
\section{Bound on anomalous energy flow}
\label{app:bound on REF}

For the convenience of the reader, in this section we re-propose a derivation for the bound on the anomalous energy flow for two interacting systems in the presence of correlations. Similar bounds appeared previously, e.g., in Refs. \cite{Jennings_2010,Reeb_2014,micadei2019}. To derive the bound, we first remind the definition of the (quantum) relative entropy \cite{umegaki1954conditional}, i.e.
\begin{align}
    S(\varrho\|\sigma) := \Tr[\varrho (\log \varrho - \log \sigma)].
\end{align}
Moreover, the von Neuman entropy is defined as
\begin{align}
    S(\varrho) := -\Tr[\varrho \log \varrho].
\end{align}
Finally, the quantum mutual information is defined as
\begin{align}
    I(A:B)_{\varrho_{\s{AB}}} := S(\varrho_{\s{AB}} \| \varrho_{\s{A}} \ot \varrho_{\s{B}}).
\end{align}
Consider a bipartite system $\varrho_{\s{AB}}$ with thermal marginals, i.e. $\varrho_{\s{A}} := \Tr_{\s{B}}[\varrho_{\s{AB}}] = \gamma_{\s{A}}$ and $\varrho_{\s{B}} := \Tr_{\s{A}}[\varrho_{\s{AB}}] = \gamma_{\s{B}}$, undergoing an entropy non-decreasing quantum channel $\mathcal{E}$, i.e. {a completely-positive (CP) trace-preserving (TP) linear map, that is}
\begin{align}
    \varrho_{\s{AB}} \rightarrow \mathcal{E}(\varrho_{\s{AB}}) = \sigma_{\s{AB}}, \,\,\, \text{where}\,\,\, S(\varrho_{\s{AB}}) \leq S(\sigma_{\s{AB}}).
\end{align}
For the initial and the evolved state of system $\s{A}$ we can write
\begin{align}
    \label{eq:app_entropy_sigma_gamma}
    S(\sigma_{\s{A}}||\gamma_{\s{A}}) &= -S(\sigma_{\s{A}}) - \textup{Tr}\left[\sigma_{\s{A}}\log\gamma_{\s{A}}\right] \\ \label{eq:app_entr}
  &= -S(\sigma_{\s{A}}) + \beta_{\s{A}}\textup{Tr}\left[\sigma_{\s{A}} H_{\s{A}} \right] + \log \mathcal{Z}_{\s{A}}.
\end{align}
From the definition of relative entropy it follows that $S(\gamma_{\s{A}}||\gamma_{\s{A}}) = 0$. Consequently, we can write
\begin{align}
\nonumber   S(\sigma_{\s{A}}||\gamma_{\s{A}}) &=   S(\sigma_{\s{A}}||\gamma_{\s{A}}) - S(\gamma_{\s{A}}||\gamma_{\s{A}}) \\
\nonumber   &\begin{tabular}{l}
                $=-S(\sigma_{\s{A}}) + \beta_{\s{A}}\textup{Tr}\left[\sigma_{\s{A}} H_{\s{A}} \right] + \log\mathcal{Z}_{\s{A}}$\\
                $\;\;\;\;\;\;+ S(\gamma_{\s{A}}) -  \beta_{\s{A}} \textup{Tr}\left[\gamma_{\s{A}} H_{\s{A}} \right] - \log\mathcal{Z}_{\s{A}}$
                \end{tabular} \\
&=\beta_{\s{A}}\Delta E_{\s{A}} - \Delta S_{\s{A}},
\label{eq:rel_ent_energy}
\end{align}
where the term $\Delta  S_{\s{A}} =  S(\sigma_{\s{A}})-S(\gamma_{\s{A}}) $ represents the change in the entropy of system $\s{A}$.

In general, we define the change of the total energy of system AB as $W := \Delta E_{\s{A}}+\Delta E_\s{B} $, and hence $\Delta E_{\s{A}} = W - \Delta E_\s{B} $. If $\mathcal{E}$ is a unitary operation, then $W$ represents the total work exchange. When $\mathcal{E}$ is a non-unitary process, then $W$ is the total energy exchange composed of two contributions: the work and the heat exchanges.

Let us now write Eq. (\ref{eq:rel_ent_energy}) separately for both subsystems $\s{A}$ and $\s{B}$ and add them up. This gives us a formula for the energy exchange between two systems under an arbitrary quantum channel, i.e.
\begin{align}
    (\beta_{\s{A}}-\beta_{\s{B}})\, \Delta E_{\s{A}}  = \Delta S_{\s{A}}+\Delta S_{\s{B}}  -\beta_{\s{B}}\,W  + \sum_{i = A,B} S(\varrho_i||\gamma_i).\label{eq: change in energy of A}
\end{align}

Since $\sum_i S(\varrho_i||\gamma_i)\geq0$, we thus have the inequality 
\begin{align}
   (\beta_{\s{A}}-\beta_{\s{B}})\, \Delta E_{\s{A}} \geq \Delta S_{\s{A}}+\Delta S_{\s{B}} - \beta_{\s{B}}\,W.
    \label{eq:gen energy entropy ineq}
\end{align}

Both types of dynamics of the system $\s{AB}$ described in this work (i.e. unitary and unitary with a catalyst) are entropy non-decreasing processes, i.e. they satisfy 
\begin{equation}
    \Delta S_{\s{AB}}=S\left(\sigma_{\s{AB}}\right)-\, S\left(\varrho_{\s{AB}}\right) \geqslant 0.
    \label{eq:entropy difference ineq}
\end{equation}
Combining inequalities \eqref{eq:gen energy entropy ineq} and \eqref{eq:entropy difference ineq} and using the fact that $\Delta \mathcal{I}=\Delta S_{\s{A}}+\,\Delta S_{\s{B}}-\,\Delta S_{\s{AB}}$, one can obtain 
\begin{align}
   (\beta_{\s{A}}-\beta_{\s{B}})\, \Delta E_{\s{A}} \geq \Delta \mathcal{I} - \beta_{\s{B}}\,W.
    \label{eq:gen energy mutual info ineq}
\end{align}
If we now assume that $\beta_{\s{A}}\geq \beta_{\s{B}}$, then Eq.\,\eqref{eq:gen energy entropy ineq} tells us that extracting energy from the cold system is possible by either exploiting the initial correlations or investing external energy on the total system.

\section{Role of non-degenerate coherences in energy preserving dynamics}\label{sec: use of nondegenerat coherences} 
Here we demonstrate that energy coherences outside of the degenerate energy subspaces do not affect the flow of energy in the regime of energy-preserving dynamics. To do so, we begin by examining the state's decomposition in Eq.~(\ref{eq: Blocky density operator}), i.e.
\begin{align}
    \varrho_{\s{AB}} &= \sum_{\nu = 1}^M \varrho_{\s{AB}}^{\nu\nu} + \sum_{\nu \neq \mu}^M \varrho_{\s{AB}}^{\nu\mu}.   
   \label{eq:app_block_dm}
    \end{align} 
The terms $\varrho_{\s{AB}}^{\nu \mu}$ for $\nu \neq \mu$ describe all energy coherences outside of the degenerate subspaces. We are going to show that they do not contribute to the average (local) energy change $\Delta E_{\s{A}}$. Let us define
\begin{align}
    E_{\s{A}}^{\nu \mu} &:= \Tr[ \left(H_{\s{A}}\otimes \mathds{1}_\s{B}\right) \,\varrho_{\s{AB}}^{\nu\mu}\,  ],\\
    \widetilde{E}_{\s{A}}^{\nu \mu} &:= \Tr[ \left(H_{\s{A}}\otimes \mathds{1}_\s{B}\right) \,U\varrho_{\s{AB}}^{\nu\mu}U^{\dagger}  ].
\end{align}
With the above definitions we have $\Delta E_{\s{A}} = \sum_{\nu, \mu} (\widetilde{E}_{\s{A}}^{\nu \mu} - {E}_{\s{A}}^{\nu \mu})$.
The contribution to the initial energy of subsystem $\s{A}$ corresponding to $\varrho_{\s{AB}}^{\nu \mu}$ is given by
\begin{align}
    E_{\s{A}}^{\nu \mu} &= \s{Tr}[ \left(H_{\s{A}}\otimes \mathds{1}_\s{B}\right) \,\varrho_{\s{AB}}^{\nu\mu}\,  ]\\
    & =\s{Tr}\left[\Pi_\mu \, \left(H_{\s{A}}\otimes\mathds{1}_\s{B}\right)\, \Pi_\nu  \,\varrho_{\s{AB}}  \right] \\
    &= \delta_{\nu \mu} E_{\s{A}}^{\nu \nu},
    \label{eq: initial eneergy off-diag appendix}
\end{align}
where we used the fact that $H_{\s{A}}\otimes\mathds{1}_\s{B}$, $\Pi_\mu$ and $\Pi_\nu$ have the same block-diagonal structure.

We now compute the contribution of the term $\varrho_{\s{AB}}^{\nu \mu}$ to the final energy of system $\s{A}$, i.e.
\begin{align}
   \widetilde{E}_{\s{AB}}^{\nu \mu}&= \s{Tr}[ (H_{\s{A}}\otimes \mathds{1}_\s{B})\, U_{\s{AB}}   \,\varrho_{\s{AB}}^{\nu\mu}\, U_{\s{AB}}^\dagger  ]\\
    &= \s{Tr}[(H_{\s{A}}\otimes \mathds{1}_\s{B})\prod_{i = 1}^M U_{i}  \, \left(\Pi_\nu \varrho_{\s{AB}} \,\Pi_\mu \right)\prod_{j = 1}^M U_{j}^\dagger ]\\
    & = \s{Tr}\left[\Pi_\mu U_\mu^\dagger \, \left(H_{\s{A}}\otimes\mathds{1}_\s{B}\right)\, U_\nu \Pi_\nu \,\varrho_{\s{AB}}  \right]. \label{eq: using cyclic prop of trace appendix}
\end{align}
In the second line we used Eq. (\ref{eq: block diagonal preserving unit}), i.e. the fact that an energy-preserving unitary $U$ is block-diagonal in the energy basis, as well as $\varrho^{\nu\mu}_{\s{AB}} = \Pi_{\nu} \varrho_{\s{AB}}\Pi_{\mu}$. In the last line we used the cyclic property of the trace and the fact that  $\prod_i U_i \Pi_j = U_j\Pi_j=\Pi_j U_j$. Since the operators $H_{\s{A}}\otimes\mathds{1}_\s{B}$, $U_\mu$ and $U_\nu$ have the same block-diagonal form (in the energy-basis), we have that $\Pi_\mu U_\mu U_\nu\Pi_\nu=\, \delta_{\mu\nu}\openone_{\nu}$, it is straightforward to show that 
\begin{equation}
   \Pi_\mu U_\mu^\dagger \, \left(H_{\s{A}}\otimes\mathds{1}_\s{B}\right)\, U_\nu\Pi_\nu=\, \delta_{\mu\nu} \Pi_\mu U_\mu^\dagger \, \left(H_{\s{A}}\otimes\mathds{1}_\s{B}\right)\, U_\mu\Pi_\mu.
\end{equation}

Therefore $\widetilde{E}_{\s{AB}}^{\nu \mu} = \delta_{\nu \mu} \widetilde{E}_{\s{AB}}^{\nu \nu}$ and we can infer that $\Delta E_{\s{A}} = \sum_{\nu} (\widetilde{E}^{\nu \nu}_{\s{AB}} - E^{\nu \nu}_{\s{AB}})$. Consequently,  energy coherences outside of the blocks of degenerate energy do not contribute to the energy flow in the process. Naturally, a similar derivation can be carried out for subystem $\s{B}$, leading to the
same result, i.e. $\widetilde{E}_{\s{B}}^{\mu\nu} =   {E_{\s{B}}^{\mu\nu}}=0 $ for $\mu\neq \nu$.

\section{Proof of \textbf{Theorem 1}}
\label{app:proof_theorem1}
In this section we prove Theorem \ref{thm1}. For that, we 
construct a unitary $U_{\star}$ capable of optimally cooling subsystem $\s{A}$ among all energy-preserving unitaries. We then use this unitary to find the final state $\sigma^{\star}_{\s{AB}}$ i.e. a state in which the optimal amount of energy has been transferred from $\s{A}$ to $\s{B}$. 

To this end, we consider the block decomposition of the density operator $\varrho_{\s{AB}}$ presented in Eq.~(\ref{eq: Blocky density operator}). Our goal is to determine the unitary $U_{\star}$ which leads to the optimal value of the optimization problem from Eq. (\ref{eq:energy_flow_rev}). First, observe that any unitary $U$ which satisfies $[U, H_{\s{AB}}] = 0$ only acts on the degenerate subspaces of $\s{AB}$. Therefore, the terms $\varrho^{\nu \mu}_{\s{AB}}$ for $\nu \neq \mu$ 
in Eq.~(\ref{eq: Blocky density operator}) do not contribute to $\Delta E_{\s{A}}$  (see Appendix~\ref{sec: use of nondegenerat coherences}), and can thus be ignored in subsequent calculations. In other words, for (non-catalytic) energy-preserving dynamics, correlations described by terms outside of the degenerate energy subspaces do not affect energy flows between subsystems. 

Due to the above considerations we only need to consider a block-diagonal density operator of the form $\varrho^{\text{BLOCK}}_\s{AB} = \sum_{\nu = 1}^M \varrho^{\nu\nu}_\s{AB}$, where each block $\varrho^{\nu\nu}_\s{AB}$  is processed using an independent unitary acting within the degenerate energy subspace $\nu$. To determine such a unitary we will use the approach discussed in Ref.~\cite{Exponential_Improve_for_Quant_Cool}, see also Refs.~\cite{Allahverdyan_2004,Afsary_2020}. More specifically, notice that for each degenerate energy subspace $\nu$ we can write
\begin{align}
    \Pi_v \varrho_{\s{AB}} \Pi_v = \varrho_{\s{AB}}^{\nu \nu} = \sum_{n = 1}^{m_{\nu}} p_n^{\nu \downarrow} \dyad{\phi_{n}^{\nu \downarrow}},
\end{align}
where $\{p_n^{\nu \downarrow}\}$ are eigenvalues of $\varrho^{\nu\nu}_{\s{AB}}$ arranged non-increasingly \emph{within} the energy subspace $\nu$ and $\ket{\phi_{n}^{\nu \downarrow}}$ are the associated eigenvectors. Applying an energy conserving unitary $ U= \prod_{\nu=1}^{M} U_{\nu}$ so that $\sigma_{\s{AB}} = U \varrho_{\s{AB}} U^{\dagger}$ and $U_{\nu} = (u^{\nu}_{nm})_{n,m}$, leads to the average local energy on $\s{A}$ given by
\begin{align}
    E_{\s{A}}(\sigma_{\s{A}}) = \sum_{\nu, n,m} |u_{nm}^{\nu}|^2 p_{n}^{\nu \downarrow} \varepsilon_{m}^{\s{A}}
\end{align}
The optimal unitary that leads to the minimal energy on the subsystem $\s{A}$ performs two actions: first diagonalizes a given energy subpsace in the energy basis and then reorders the eigenvalues to minimize the energy of $\s{A}$. In our current notation this means that $u_{nm}^{\nu} = \delta_{nm}$ for every $\nu$.

 Let us now decompose the projectors $\Pi_{\nu}$ onto the degenerate energy subspaces as $\Pi_v = \sum_{n=1}^{m_{\nu}} \dyad{\mathcal{E}_n^{\nu}}$, where the vectors $\{\ket{\mathcal{E}^{\nu}_{n}}\}$ are any vectors that form the basis of degenerate subspace $\nu$ which are sorted in the order of non-increasing local energies of $\s{A}$, i.e. such that
\begin{align}
\braket{\mathcal{E}_{n}^{\nu}|H_{\s{A}} \ot \mathbb{1}_{\s{B}}|\mathcal{E}_{n}^{\nu}} \geq \braket{\mathcal{E}_{k}^{\nu}|H_{\s{A}} \ot \mathbb{1}_{\s{B}}|\mathcal{E}_{k}^{\nu}} \quad \text{for $n \geq k$}.
\end{align}
With this notation the energy-conserving unitary that minimizes $\Delta E_{\s{A}}$ takes the form
\begin{align}
  U &= \prod_{\nu=1}^{M} U_{\nu}, \quad \text{where} \quad U_{\nu} :=  \sum_{n=1}^{m_\nu}
  \ketbra{\mathcal{E}^{\nu}_{n}}{\phi^{\nu\downarrow}_{n}}+\sum_{\mu\neq\nu}\Pi_\mu,
  \label{eq: Optimal unitary energy conserving}
\end{align}
Let us write the state $\varrho^{\text{BLOCK}}_\s{AB}$ in the basis of $\{\ket{\phi^{\nu}_n}\}$, i.e.
\begin{align}
   \varrho^{\text{BLOCK}}_\s{AB} = \sum_{\nu = 1}^M \varrho^{\nu\nu}_\s{AB} &= \sum_{\nu=1}^M \left(\sum_{n=1}^{m_\nu} p_{n}^{\downarrow} \ket{\phi^{\nu \downarrow}_{n}}\bra{\phi^{\nu \downarrow}_{n}}\right).
\end{align}

Applying the unitary from Eq.~(\ref{eq: Optimal unitary energy conserving}) to the above state yields
\begin{align}
    \sigma_\s{AB}^{\star} &= \prod_{\nu=1}^M U_\nu \left(\sum_{n=1}^{m_\nu} p_{n}^{\nu \downarrow} \ket{\phi^{\nu \downarrow}_{n}}\bra{\phi^{\nu \downarrow}_{n}}       \right)U_\nu^{\dagger}\\
    &= \sum_{\nu=1}^M\left (  \sum_{n=1}^{m_\nu}{p}^{\nu \downarrow}_{n} \ketbra{\mathcal{E}^{\nu}_{n}}  \right).
\end{align}
The reduced state on $\s{A}$ therefore becomes $\sigma_{\s{A}}^{\star}:= \Tr_{\s{B}}\left[ \sigma^{\star}_{ \s{AB} }\right]$. In order to determine its energy it is enough to look at the energy occupations, i.e. $q_i^{\star} = \langle \varepsilon^{\s{A}}_i |\sigma_{\s{A}}| \varepsilon^{\s{A}}_i \rangle$. They are given by
\begin{align}
    q_i^{\star} &= \Tr[\dyad{\varepsilon_i^{\s{A}}} \ot \mathbb{1}_{\s{B}} \sigma_{\s{AB}}^{\star}] \\
    &= \sum_{\nu=1}^M \sum_{n=1}^{m_\nu}{p}^{\nu \downarrow}_{n} \Tr[\dyad{\varepsilon_i^{\s{A}}} \ot \mathbb{1}_{\s{B}} \dyad{\mathcal{E}_n^{\nu}}] \\
    &= \sum_{\nu=1}^M \sum_{n=1}^{m_\nu}{p}^{\nu \downarrow}_{n} \braket{\mathcal{E}_n^{\nu}| \left(\dyad{\varepsilon_i^{\s{A}}}\ot \mathbb{1}_{\s{B}}\right)|\mathcal{E}_n^{\nu}}.
\end{align}
This completes the proof of the theorem. 

\section{Fundamental bound on catalytic AEF}
\label{app:bound_on_REF}

In the main text we have shown that, for a bipartite state, the energy flow can be enhanced when using appropriately tuned states that act as catalysts. It is natural to ask what is the maximal energy flow that can be obtained with the help of arbitrary catalysts. 

Let us now consider system $\s{AB}$ and arbitrary ancillary system $\s{C}$, which acts as a catalyst, i.e. remains invariant at the end of the energy-conserving unitary processes. The quantity that captures this can be formally defined in the following way

\begin{align}
    \label{eq:catalytic_reversal_arbU_explicit}
    \Delta E_{\s{A}}^{\star , \s{c}} := \hspace{10pt} \min_{U, \omega_\s{C}}& \hspace{10pt} E(\sigma_{\s{A}}) - E(\varrho_{\s{A}}) \\
                \s{subject\,\, to} & \hspace{10pt} U(\varrho_\s{AB} \ot \omega_\s{C})U^{\dagger} = \sigma_\s{ABC} \\
                & \hspace{10pt} \sigma_\s{C} = \omega_\s{C},
                \hspace{4pt} \omega_\s{C} \geq 0, \hspace{4pt}\Tr[\omega_\s{C}] = 1, \\
    & \hspace{10pt} U U^{\dagger} = U^{\dagger} U = \mathbb{1}_\s{ABC},  \\
    & \hspace{10pt} [U, H_\s{ABC}] = 0.
\end{align}
Unfortunately, computing $\Delta E_{\s{A}}^{\star , \s{c}}$ explicitly is a difficult problem. In what follows we derive nontrivial lower bounds on $\Delta E_{\s{A}}^{\star , \s{c}}$.  For that, we make use of the \emph{Catalytic Entropy Theorem} which provides the necessary and sufficient conditions for transforming general states using \emph{arbitrary} unitary operations in the presence of \emph{arbitrary} catalysts \cite{Wilming_2021}. The theorem states that for any two density operators $\varrho_{\s{AB}}$ and $\sigma_{\s{AB}}$, there exists a density matrix $\omega_{\s{C}}$ and a unitary $U$ such that $\Tr_{\s{C}}[U(\varrho_{\s{AB}} \ot \omega_{\s{C}})U^{\dagger}] = \sigma_{\s{AB}}$ and $\Tr_{\s{AB}}[U(\varrho_{\s{AB}} \ot \omega_{\s{C}})U^{\dagger}] = \omega_{\s{C}}$ if and only if $S(\sigma_{\s{AB}}) \geq S(\varrho_{\s{AB}})$. This makes it clear that, instead of searching for optimal unitaries $U$ and density matrices $\omega_{\s{C}}$ in the optimization problem (\ref{eq:catalytic_reversal_arbU_explicit}), it is enough to search over all density matrices $\sigma_{\s{AB}}$ on the system $\s{AB}$ with von Neumann entropy bounded from below by $S(\varrho_{\s{AB}})$. Still, the unitaries that we consider here are arbitrary, i.e. our solutions will not respect the constraint $[U, H_\s{ABC}] = 0$ and therefore can, in general, achieve a lower value in the optimization problem (\ref{eq:catalytic_reversal_arbU_explicit}). In order to make our bound tighter we add a linear constraint on the average energy, i.e. $\Tr[H_\s{AB}\sigma_\s{AB}] = \Tr[H_\s{AB}\varrho_\s{AB}]$. This allows us to lower bound the optimal value of the optimization problem (\ref{eq:catalytic_reversal_arbU_explicit}) using the following optimization problem:
\begin{align}
    \label{eq:simplified_catalytic_sdp}
    \Delta E_{\s{A}}^{(\s{bound})} := \hspace{20pt} \min_{\sigma_{\s{AB}}}& \hspace{10pt}  E(\sigma_{\s{A}}) - E(\varrho_{\s{A}}) \\
                \s{subject\,\, to} & \hspace{10pt} S(\sigma_{\s{AB}}) \geq S(\varrho_{\s{AB}}) \\
                & \hspace{10pt} E(\sigma_{\s{AB}}) = E(\varrho_{\s{AB}})  
\end{align}
Remarkably, the above problem is an instance of a semi-definite program (SDP), which implies that it can be solved
efficiently using existing numerical techniques. 


\section{Toy example of AEF activation}
\label{app:toy_example}
In this Appendix we present a simple toy example that demonstrates the role of catalysts in generating desired flows of energy in classically-correlated systems. Consider a bipartite system prepared in a classically-correlated state
\begin{align}
    \varrho_\s{AB} = \frac{1}{2} \left(\dyad{00} + \dyad{11}\right),
\end{align}
with local Hamiltonians $H_x = \varepsilon \dyad{1}_{x}$ for $x \in \{\s{A},\s{B}\}$. The local marginals of the above state are given by  $\Tr_{\s{A}}[\varrho_\s{AB}] = \gamma_\s{B}$  and $\Tr_{\s{B}} [\varrho_{\s{AB}}] = \gamma_\s{A}$, where $\gamma_\s{A} = \gamma_\s{B}$ are thermal states at equal temperatures $T_\s{A} = T_\s{B} = \infty$. Since the degenerate energy subspace of $\varrho_{\s{AB}}$ is not populated, there is no unitary $U$ that satisfies $[H_\s{A} + H_\s{B}, U] = 0$ and which could lead to a flow of energy between subsystems $\s{A}$ and $\s{B}$.

Let us now add a third system $\s{C}$ (a catalyst) with Hamiltonian $H_{\s{C}} = \varepsilon \dyad{1}_{\s{C}}$ prepared in a maximally-mixed state $\omega_{\s{C}} = \frac{1}{2} \mathbb{1}_{{\s{C}}}$. Let the global system evolve according to an energy-conserving unitary $U$. The action of $U$ on the energy basis of the Hilbert space of three qubits $\ket{ijk} := \ket{i}_\s{A} \ot \ket{j}_\s{B} \ot \ket{k}_\s{C}$ for $i,j,k \in \{0,1\}$ is given by
\begin{align}
    U \ket{001}_{} &= \ket{010}_{},  &U \ket{110}_{} = \ket{011}_{}, \\
    U \ket{010}_{} &= \ket{001}_{},  &U \ket{011}_{} = \ket{101}_{}, \\
    U \ket{101}_{} &= \ket{110}_{},
\end{align}
and $U$ furthermore acts as identity on the remaining elements of the energy basis. It can be easily verified that $U$ satisfies $[U, H_\s{A}+H_\s{B}+H_\s{C}] = 0$ and evolves the 
 composite system into  $\sigma_\s{ABC}:= U(\varrho_\s{AB} \ot \omega_\s{C})U^{\dagger}$, where the reduced states become
\begin{align}
    \sigma_\s{A} = \frac{3}{4} \dyad{0} + \frac{1}{4} \dyad{1}, \quad \sigma_\s{B} = \frac{1}{4} \dyad{0} + \frac{3}{4} \dyad{1},\!
\end{align}
and furthermore $\sigma_\s{C} = \omega_\s{C}$. The energy change on $\s{A}$ is given by $\Delta E_{\s{A}} = E(\sigma_{\s{A}}) - E(\varrho_{\s{A}}) = -1/4 < 0$. That is, the system $\s{A}$ becomes colder and the system $\s{B}$ becomes hotter. This demonstrates that in the scenario with the energy-conservation law catalysts can enable anomalous flow of energy even for fully classically-correlated systems. Indeed, both systems $\s{A}$ and $\s{B}$ are incoherent and classically-correlated states, and the catalyst $\s{C}$ is incoherent in the energy eigenbasis.

\bibliographystyle{apsrev4-2}
\bibliography{references}

\end{document}